\documentclass{article} 
\usepackage{colm}
\usepackage[colorlinks = true,
            linkcolor = blue,
            urlcolor  = blue,
            citecolor = blue,
            anchorcolor = blue]{hyperref}   
\usepackage{microtype}
\usepackage{hyperref}
\usepackage{url}
\usepackage{booktabs}
\usepackage{enumitem}
\usepackage{multicol}
\usepackage{multirow}
\usepackage{CJKutf8}
\usepackage{amsmath}
\usepackage{siunitx}
\usepackage{floatflt}
\usepackage{wrapfig}
\usepackage{listings}
\usepackage{amsfonts}
\usepackage{CJKutf8}
\usepackage{booktabs}
\usepackage{tabularx}
\usepackage{graphicx}
\usepackage{algorithm}
\usepackage{algpseudocode}
\usepackage{threeparttable}

\usepackage{caption}

\newcommand\blfootnote[1]{%
  \begingroup
  \renewcommand\thefootnote{}\footnote{#1}%
  \addtocounter{footnote}{-1}%
  \endgroup
}

\title{Level-Navi Agent: A Framework and benchmark for Chinese Web Search Agents}

\author{Chuanrui Hu$^{1*}$$^{\dagger}$, Shichong Xie$^{2*a}$, Baoxin Wang$^{1}$, \\ \textbf{Bin Chen$^{1}$, Xiaofeng Cong$^{1}$, Jun Zhang$^{2\ddagger}$}\\
\\
$^1$ 360 AI Research \\
$^2$ School of Artificial Intelligence, Anhui University}

\colmfinalcopy 

\begin{document}

\maketitle

\blfootnote{$^*$ Equal Contribution. $^{a}$ Intern at 360 AI Research. $^{\dagger}$ Project Lead. $^{\ddagger}$ Corresponding author}

\begin{abstract}
Large language models (LLMs), adopted to understand human language, drive the development of artificial intelligence (AI) web search agents. Compared to traditional search engines, LLM-powered AI search agents are capable of understanding and responding to complex queries with greater depth, enabling more accurate operations and better context recognition. However, little attention and effort has been paid to the Chinese web search, which results in that the capabilities of open-source models have not been uniformly and fairly evaluated. The difficulty lies in lacking three aspects: an unified agent framework, an accurately labeled dataset, and a suitable evaluation metric. To address these issues, we propose a general-purpose and training-free web search agent by level-aware navigation,  Level-Navi Agent, accompanied by a well-annotated dataset (Web24) and a suitable evaluation metric. Level-Navi Agent can think through complex user questions and conduct searches across various levels on the internet to gather information for questions. Meanwhile, we provide a comprehensive evaluation of state-of-the-art LLMs under fair settings. To further facilitate future research, source code is available at \href{https://github.com/chuanruihu/Level-Navi-Agent-Search}{Github}.
\end{abstract}
\section{Introduction}
Information gathering is a key step in the interaction between humans and their environment. Search engines are widely used for information acquisition (\cite{BRIN1998107}). With the development of large language models (LLMs) (\cite{ye2023comprehensive}, \cite{achiam2023gpt}), AI search agents based on LLMs have become an emerging and challenging research topic \cite{Nakano2021WebGPTBQ}.

Retrieve-Augmented Generation (RAG) is used to improve the precision of model responses (\cite{ram-etal-2023-context}). Existing methods (\cite{chan2024rqrag}, \cite{siriwardhana2023improving}) leverage the powerful language capabilities of LLMs to perform retrieval based on user queries and use the retrieved relevant texts to improve the reliability of the model's answers. This advanced ability to understand and analyze questions exceeds that of traditional search engines, driving a revolutionary transformation in AI-powered search (\cite{spatharioti2023comparing}). However, these methods do not further explore how LLMs handle complex questions. The simple text retrieval approach cannot fully align with web search scenarios. And irrelevant texts retrieved have negative impacts on the quality of responses (\cite{asai2023self}).

\begin{figure}[hb!]
    \centering    
    \includegraphics[width=0.65\columnwidth]{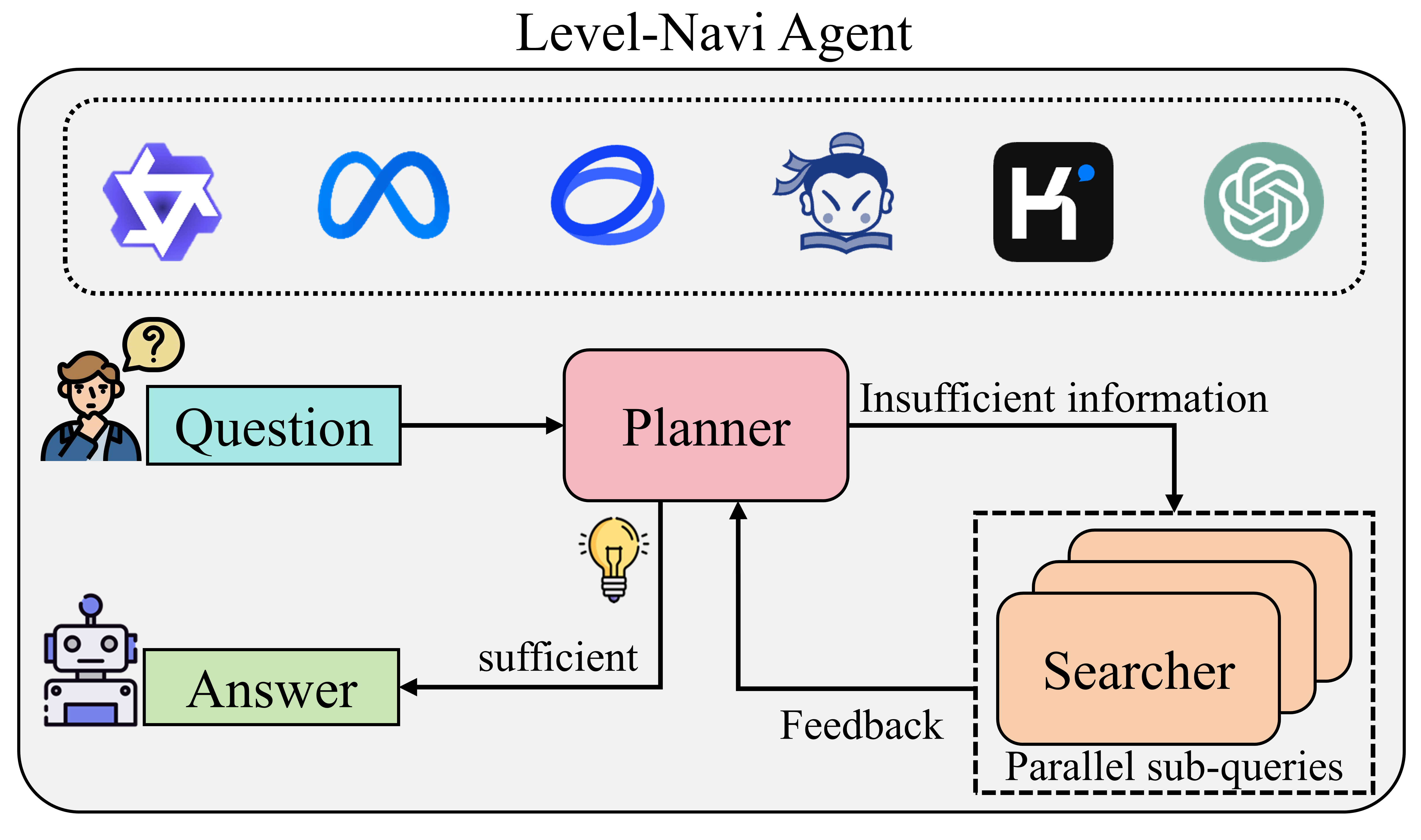}
    \caption{Pipeline of Our Level-Navi Agent.}
    \label{fig:first}
\end{figure}

Therefore, refined methods(~\cite{chen2024mindsearch}, \cite{Reddy2023SmartBookAS}) are proposed to construct AI search agents. Mindsearch(~\cite{chen2024mindsearch}) employs the concept of Directed Acyclic Graphs to structure the agent's plan, breaking down complex reasoning questions with the aim of simulating the human mind, thereby striving to deliver more comprehensive answers. Infogent(~\cite{reddy2024infogent}) utilizes an information aggregation approach to update the retrieved information. Determine whether the retrieved texts meet the required conditions and improve the accuracy of responses by controlling the quality of the information. These methods achieve promising results under detailed process planning. However, the research community still lacks comprehensive studies that can genuinely reveal the true capabilities of various open-source and closed-source LLMs in the web search scenario.

Existing LLM-driven search agents require fine-tuning or rely on high-performance close-source models, making it difficult for researchers to investigate the capabilities of various LLMs due to their costs. Datasets for evaluating the capabilities of LLMs in Chinese scenarios are constructed, such as CMMLU (\cite{li2023cmmlu}) and AlignBench (\cite{liu-etal-2024-alignbench}). The advent of these datasets shows the demand for model evaluation in real-world Chinese contexts. However, in the field of web search, a suitable Chinese web search dataset for quantitative evaluation is lacking. Meanwhile, we reveal that traditional metrics like F1 and ROUGE(~\cite{lin2004rouge}) do not consider the semantic information across various versions, which poses challenges when comparing the performance of different models.

To address the aforementioned issues, we propose a training-free AI search agent framework for both open-source and close-source models, as illustrated in Fig. \ref{fig:first}. Meanwhile, we provide a new Chinese web search dataset and a new evaluation metric to evaluate the performance of LLMs in the Chinese task. Overall, our contributions are as follows.
\begin{itemize}
    \item \textbf{We propose a general-purpose training-free web search agent framework, called Level-Navi Agent.} The question from the user is first analyzed and decomposed by the Planner. Then the sub-questions are provided to the Searcher, which will collect information at different levels. By iterating this step, Level-Navi Agent eventually collects enough information to answer the initial question. Level-Navi Agent does not require training, allowing any open-source LLM to be deployed.
    
    \item \textbf{We provide a well-annotated benchmark dataset (Web24) for Chinese web search.} Our dataset is capable of a diverse and detailed classification of questions and sources, all sourced entirely from the Chinese internet. Considering the limitations of traditional metrics, we adopted four reasonable metrics to evaluate the ability of different LLMs when execute the Level-Navi Agent. Through our benchmark, the performance of different LLMs for AI web search is clearly presented. 
    
    \item \textbf{We reveal the factors that limit model performance in executing Chinese web search agent tasks.} First, we find that the model exhibits an ``overconfidence" phenomenon, where it refrains from calling functions for web searches even when it does not know the answer, leading to incorrect responses. Second, the model demonstrates low ``task fidelity" during task execution, meaning it fails to fully understand our instructions, resulting in non-compliant answers and poor response quality.
\end{itemize}

\section{Level-Navi Agent}\label{agent}

\begin{figure}[htbp!]
    \centering    
    \includegraphics[width=\columnwidth]{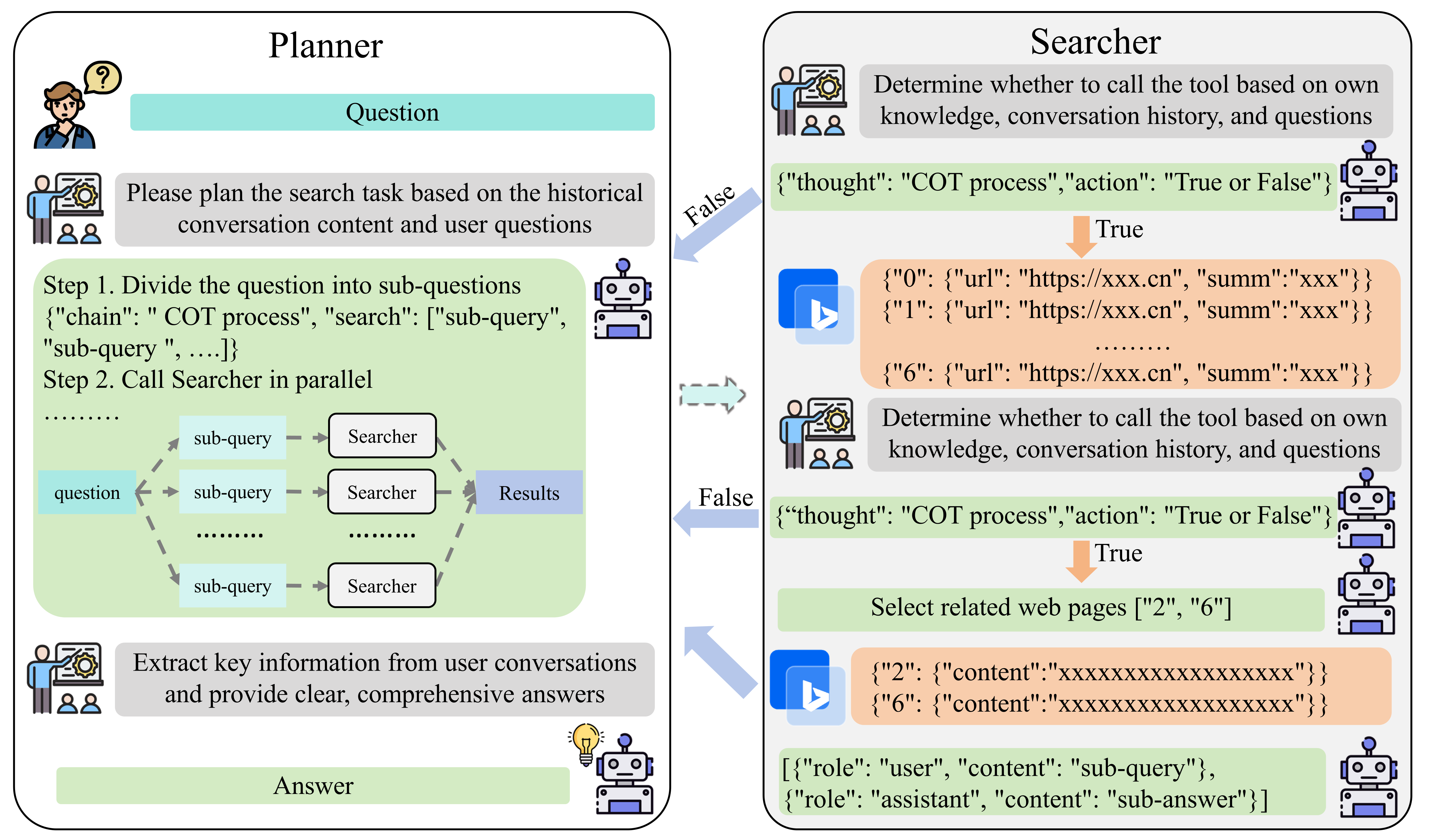}
    \caption{The framework of Level-Navi Agent.}
    \label{fig:frame}
\end{figure}

In this Section, we will introduce our Level-Navi agent. We detail the structure of our Planning Agent and Level-info Agent in Section \ref{agent1} and Section \ref{agent2}. The overall structure is shown in Fig. \ref{fig:frame}. Fig. \ref{fig:case} illustrates a real case handled by our agent.

\subsection{Planning Agent}\label{agent1}

\begin{figure}[htbp!]
    \centering    
    \includegraphics[width=\columnwidth]{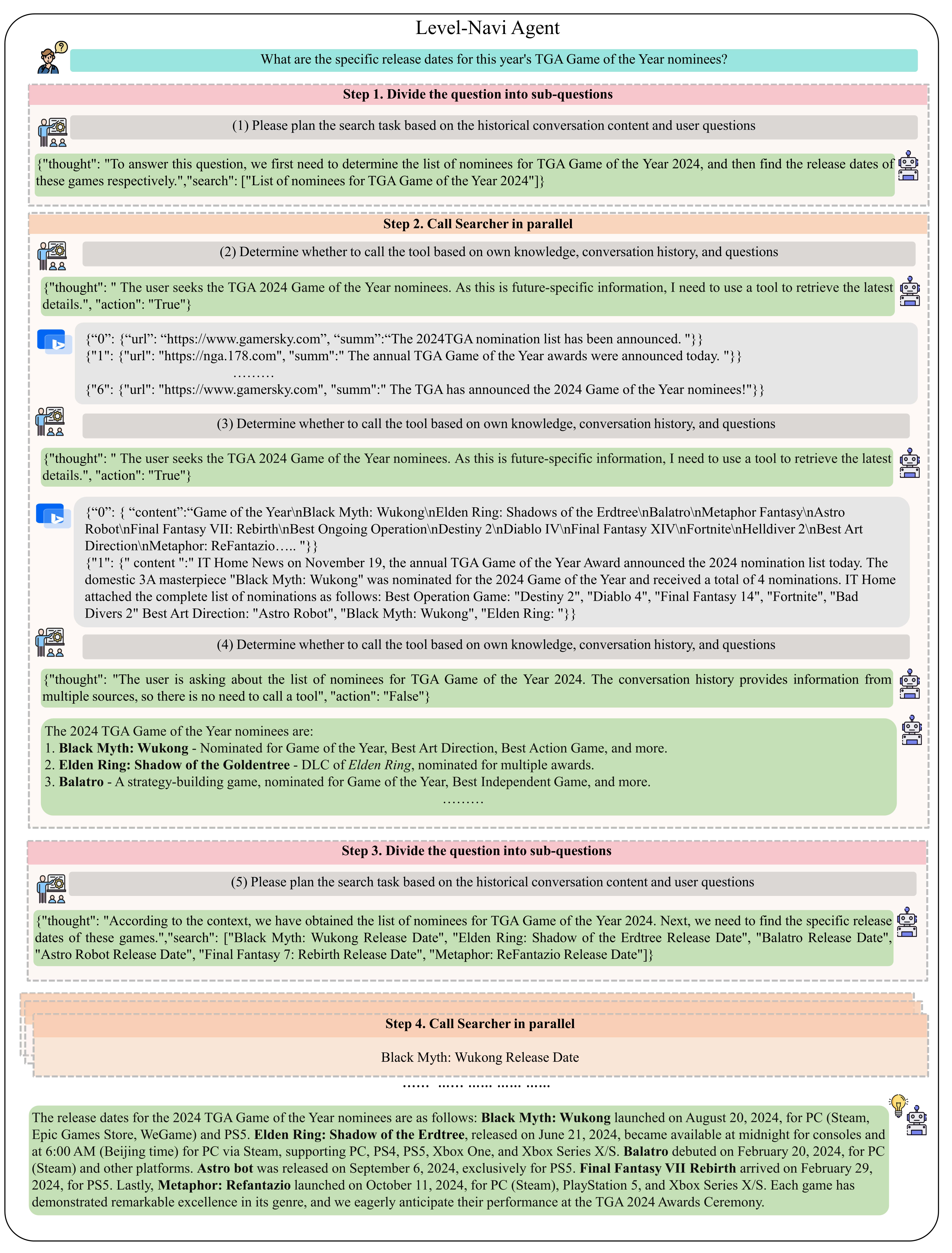}
    \caption{Our Level-Navi Agent demonstrates an example of handling a user query. The Planning Agent first requests the collection of information about nominated games. After receiving feedback, it then proceeds to search for the release date of each game in parallel (we translate the process from Chinese to English for better understanding).}
    \label{fig:case}
\end{figure}

The Planner is a key component in our design, and its structure directly affects the performance of the entire Agent to a certain extent. Next, we will provide a detailed introduction to the design of the Planning Agent. Our Planning Agent plans the trajectory path through a process of chain of thought (\cite{wei2022chain}) and iterative refinement. When user inputs a question, our Agent first understands and breaks down the problem through chain of thought. Noticing the difference from the conventional chain of thought steps (\cite{jin2024impact}, \cite{wang2023knowledge}), we do not ask the LLM to only provide a complete set of steps to solve the problem at once. We let the LLM first think through and determine the information that should be collected next, then generate a list of sub-questions that can be searched in parallel at this stage. This is because, in scenarios where complex reasoning is required, a complete plan may seem very clear; however, since we rely on another Agent to gather information, the content obtained each time is not necessarily sufficient or complete, which involves dynamically adjusting the plan every time. To avoid such a complex and redundant process design, we use prompt to enforce that the Planning Agent only giving the list of sub-question that needed to be obtained in the next step. After obtaining feedback from each step, it repeats this process iteratively until the Agent judges that the current information is sufficient to answer the question. As depicted in Fig. \ref{fig:case}, our Planning Agent has broken down the problem into multiple levels, systematically presenting the sub-problems required at each step. Through iteration, it gathers sufficient information to answer the question effectively. Algorithm \ref{algorithm:planning agent} also demonstrates the complete process of our Planning Agent.

\begin{algorithm}
\caption{Pseudocode for the Planning Agent's Process}\label{algorithm:planning agent}
\begin{algorithmic}[1] 
\Statex \textbf{Input:} $Q$ for user's question
\Statex \textbf{Output:} $R$ for agent's response
\Statex \textbf{Variables:} $H$ for a set of history context, $M$ for a list of collected information
\Statex
\State $H$ $\gets$ $\emptyset$
\State $H \cup \{Q\}$ \Comment{add question to history context}
\While{True}
    \State Result $\gets$ $CoT(H)$ \Comment{Planning Agent's thought result}
    \If{Result = "no"}
        \State $M \gets \emptyset$
        \State $Q_{sub}$ $\gets$ Result.sub-question
        \Procedure{Process}{$Q_{\text{sub}}$}
            \State \textbf{parallel for each} $q \in Q_{\text{sub}}$ \textbf{do}
                \State $M[q] \gets Searcher(q)$ \Comment{search each sub-question in parallel}
            \State \textbf{end parallel}
        \EndProcedure
        \State $H \cup \{M\}$ \Comment{add information to context}
    \Else
        \State $R$ $\gets$ Result.response
        \State \textbf{break}
    \EndIf
\EndWhile
\State \textbf{Return} $R$
\end{algorithmic}
\end{algorithm}

Our Planning Agent relies on prompt engineering, which means that our framework is a general-purpose framework without the need for fine-tuning, and any open-source or close-source model can operate within this framework. Furthermore, our agent is also more user-friendly for those with limited hardware resources, as they can flexibly deploy models of appropriate size based on their actual conditions, making it more cost-effective.

\subsection{Level-Info Agent}\label{agent2}

As depicted in the right of Fig. \ref{fig:frame}, the primary task of the Searcher is to obtain relevant information feedback to the Planner by conducting online searches based on the sub-problems received. In order to enrich the information obtained while enhancing its flexibility, we have constructed an Agent that dynamically simulates human information acquisition process through a chain of thought, which we call the Level-Info Agent.

As its name suggests, the Level-Info Agent is capable of dynamically obtaining information at various levels and can return results at any moment, significantly improving the agent's operational speed. Firstly, when faced with an input sub-query, the Level-Info Agent determines whether the information can be answered using its own knowledge. If it can, it returns the result directly; otherwise, it proceeds with a web search. Then, based on the sub-query, the Agent will call the web search function to use search engine APIs to return the results of the online search. At this step, the returned materials will only be the summary parts of the web pages. Here, we also have the Agent think and determine whether the current materials obtained can answer the sub-query. If the information is sufficient, it will provide a direct response; otherwise, it will proceed to the next step of opening relevant websites. When performing this step, we also use function calls to let the model select and open relevant websites. After obtaining the information from the web page, it will summarize and respond.

Based on the above introduction, our Level-Info Agent has up to three levels for providing information feedback. This design is inspired by human behavior in collecting information online: when faced with a question, humans may feel confident in answering it themselves. If a web search is needed, people will first browse the returned pages. If the information on the returned interface can answer the question, there is no need to open it further. The opening of web pages always occurs as the last step.

Our Level-Info Agent avoids the need to always read a large number of websites, which consumes a significant amount of tokens, and also reduces the frequency of calling search engine APIs, thereby lowering the overall cost of using the Agent.

\section{Benchmark}\label{benchmark}

In this section, we will provide a detailed introduction to the benchmark specifically designed for web search agents. The Web24 dataset will be introduced in section \ref{data composition}, and Section \ref{evaluation method} will cover our evaluation metrics.

\begin{multicols}{2}

\begin{figure}[H]
\centering
\includegraphics[width=0.9\columnwidth]{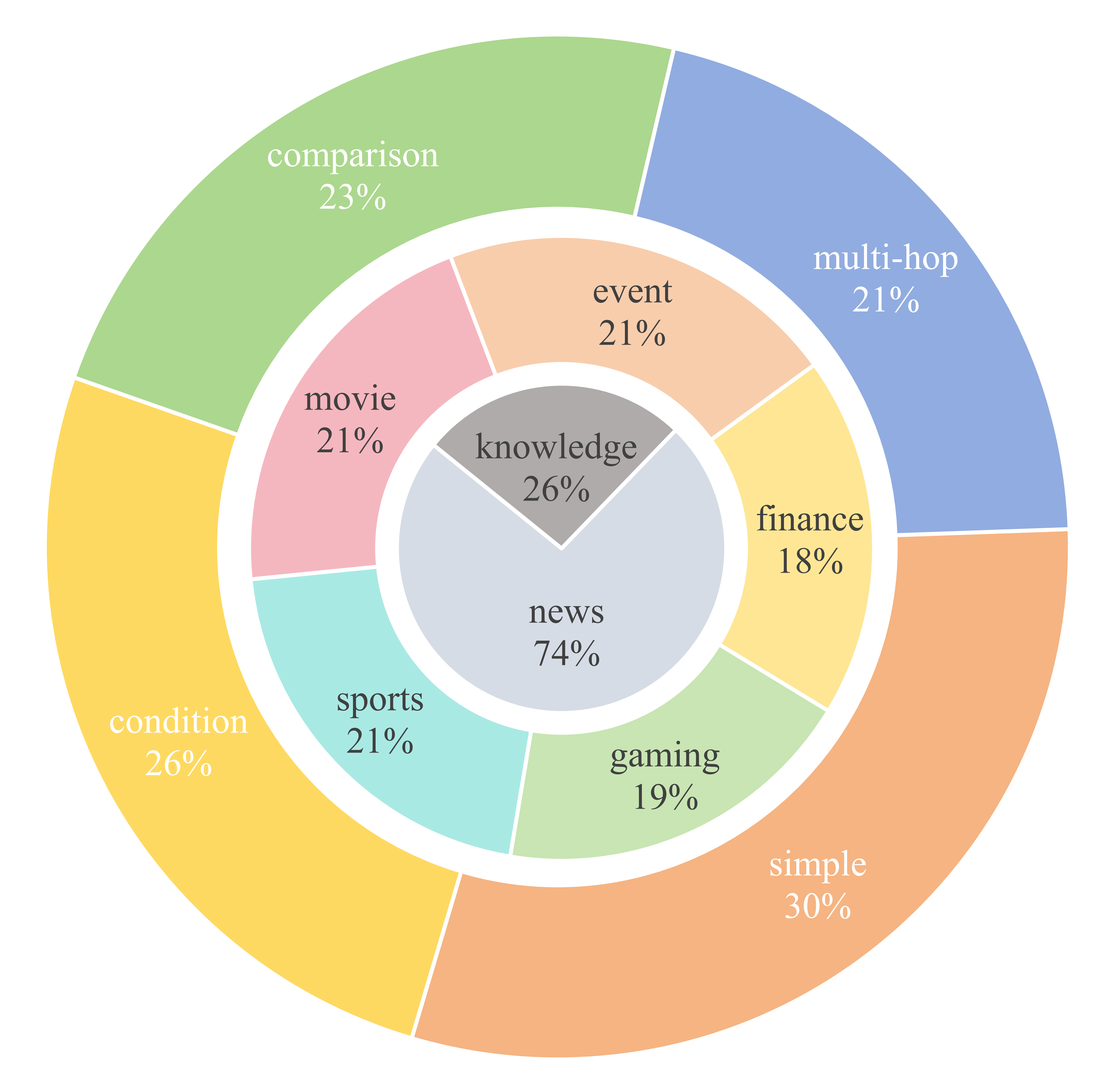}
\captionof{figure}{Source, domain and type of Web24 Dateset.}
\label{fig:data}
\end{figure}

\vspace*{0.8cm}

\centering
\begin{tabular}{lcccc|c}
\toprule
\textbf{Domain} & \textbf{S.} & \textbf{Cnd.} & \textbf{Cmp.} & \textbf{M-H} & \textbf{All} \\
\midrule
Finance & 23 & 23 & 22 & 22 & 90 \\
Gaming & 28 & 23 & 23 & 17 & 91 \\
Sports & 42 & 18 & 21 & 19 & 100 \\
Movie & 29 & 33 & 24 & 14 & 100 \\
Event & 23 & 27 & 22 & 28 & 100 \\
\midrule
All & 145 & 124 & 112 & 100 & 481 \\
\bottomrule
\end{tabular}
\captionof{table}{Distribution of Problem Domains and Types.\label{tab:domains and types}}

\end{multicols}

\subsection{Data Composition}\label{data composition}

Our Web24 dataset categorizes question-answer pairs into fine-grained divisions based on three labels: source, domain, and type.

In the process of evaluating the performance of web search agents, we aim to minimize the influence of the model's internal knowledge to genuinely assess the search capabilities. Therefore, when constructing the test dataset, we ensure that the majority of the question-answer pairs are sourced from news, as illustrated in Fig. \ref{fig:data}. All cases sourced from news are entirely from news reports on the Chinese internet before December 2024. We have provided the URLs of the report sources in the dataset for verification. For cases sourced from knowledge, our annotators cross-check to ensure that the questions are not too simple, thereby guaranteeing that the model needs to conduct a web search to obtain the answers.

To simulate as closely as possible the scenarios in which people conduct web searches in everyday life, our question-answer pairs are categorized into five domains: finance, gaming, sports, movie, and event. Each domain has been designed with specific questions and answers to reflect the information needs that users might encounter in their daily lives. 

Finally, to better express the structure of the question-answer pairs, we have categorized all question types into four categories: simple, condition, comparison, and multi-hop. The distribution of question types and domains is shown in Table \ref{tab:domains and types}. The detailed description of each type is as follows:
\begin{itemize}
    \item \textbf{Simple.} Such questions represent simple, single information requests, for example, When was the latest version of the Chinese national anthem released?
    \item \textbf{Condition.} Such questions involve constraints related to specific times or scenarios, for example, when was the announcement of the third batch of China's Time-Honored Brands?
    \item \textbf{Comparison.} Such questions involve comparing the attributes of two entities, for example, who has a higher career total points, Kobe or LeBron?
    \item \textbf{Multi-hop.} Such questions involve linking multiple pieces of information and require multi-step progressive reasoning and searching to obtain an answer, for example, where is the headquarters of the company of the courier who collected the 1500 billionth package this year?
\end{itemize}

\subsection{Evaluation Metrics}\label{evaluation method}

To comprehensively assess the capabilities of LLMs in performing web search tasks, we consider multiple aspects and use four scoring metrics to evaluate the model's capabilities holistically. Each metric takes into account the model's performance from different perspectives during the search task execution. Finally, we calculate the weighted sum of these four metrics to obtain the final assessment score.

Our detailed description of the evaluation metrics is as follows:

\textbf{Correctness Scores($S_{co}$).} In previous tasks assessing the capabilities of LLMs, the F1 metric is often used (\cite{chan2024rqrag, jiang2024mmsearch}) as an indicator to evaluate the gap between the model's responses and the ground truth answers. However, the F1 score calculated purely at the token level is not suitable for complex question-answering tasks, as semantic information and rich expressions cannot be conveyed through the F1 score, and it may even lead to misjudgments. We will discuss this issue in detail in Section \ref{traditional metrics}. To better evaluate the model response, we employ a LLM as an evaluator to assess the consistency and accuracy of the generated answers compared to the ground truth answers (\cite{yang2024crag}). We use this evaluator to score the responses on a scale of 1 to 10, and then normalize these scores to a range of 0 to 1.

\textbf{Semantic Similarity Scores($S_{simi}$).} Semantic similarity is a common method for judging text similarity. By using an embedding model (\cite{xiao2024c}), we can directly calculate the vectors of discrete tokens mapped to a high-dimensional continuous space, and directly compute the similarity between text vectors through mathematical methods. This method provides another perspective to assess the accuracy between model responses and actual answers, and this score also reflects the comprehensive ability of the model in executing web search tasks.

\textbf{Relevance Scores($S_{rele}$).} This metric primarily examines the model's faithfulness to the task execution trajectory and its ability to summarize the overall context during task execution (\cite{es-etal-2024-ragas}). The process for calculating the metric is as follows: based on the responses generated by the LLM, another evaluation LLM will generate multiple questions that are inferred from the responses, and then calculate the semantic similarity between these inferred questions and the originally given questions, then the maximum value is taken as the final score. This metric is based on a simple yet reasonable idea: if the original question can be inferred from the LLM's response, then that response should be clear and explicit. Notice that this metric does not utilize the ground truth answers, thus allowing it to focus entirely on the LLM's faithfulness to the overall task execution and its summarization capabilities.

\textbf{Searcher Count($S_{c}$).} This metric assesses the ability of LLMs to understand and break down questions. As introduced in Section \ref{agent}, the subdivided sub-questions enter the Search Agent for parallel searching, and within the Level-Info Agent, the LLM can choose to call the web search API to gather information from the internet. We have counted the number of times the Level-Info Agent is invoked in each task and use the average number of invocations as an evaluation metric. Fewer invocations indicate that the LLM has fewer sub-problems to break down, demonstrating a strong understanding capability. Additionally, fewer invocations mean that the model processes tasks more quickly and also conserves the number of web search API calls, saving costs. When calculating the total score, we use exponential decay to map the number of searcher to a score for computation.

Ultimately, we express the total score (1-100) as a weighted sum of the aforementioned four metrics. We assign the highest weight to the F1 score to highlight its importance, while semantic similarity and relevance scores are given secondary weights. The searcher count is mapped to a 1-10 scale through exponential decay and assigned the lowest weight. The formula for calculating the total score is as follows:
\begin{equation}
S_{final} = 60 \times S_{co} + 15 \times S_{simi} + 15 \times S_{rele} + 10 \times e^{-S_c}.
\end{equation}
\section{Experiments and Analysis}

In this Section, we present the experimental results on our benchmark in Section \ref{experiments result}, demonstrate the limitations of traditional methods in the experiments of Section \ref{traditional metrics}.

\begin{table}[ht]
\centering
\begin{threeparttable}
\begin{tabular}{c|c|cccccc}
\toprule
Model & Few-shot & $S_{final}$ & $S_{co}$ & $S_{rele}$ & $S_{simi}$ & $S_c$  & Pass rate\\
\midrule
\multicolumn{8}{c}{Open Source Model} \\
\hline
\midrule
\multirow{3}{*}{Internlm2.5-7B} & zero-shot & 49.48 & 0.47 & 0.81 & 0.56 & 2.62 & 0.92\\
                      & one-shot & 47.76 & 0.45 & 0.79 & 0.56 & 2.98 & 0.91\\
                      & three-shot & 49.31 & 0.47 & 0.8 & 0.56 & 2.65 & 0.95\\
\midrule
\multirow{3}{*}{Internlm2.5-20B} & zero-shot & 55.02 & 0.57 & 0.80 & 0.57 & 3.62 & 0.93\\
                      & one-shot & 57.70 & 0.61 & 0.81 & 0.58 & 3.68 & 0.96\\
                      & three-shot & 55.43 & 0.57 & 0.80 & 0.57 & 2.69 & 0.97\\
\midrule
\multirow{3}{*}{GLM-4-9B} & zero-shot & 63.25 & 0.66 & 0.83 & 0.67 & 2.16 & 0.94\\
                      & one-shot & 40.82 & 0.34 & 0.79 & 0.54 & 3.05 & 0.89\\
                      & three-shot & 43.43 & 0.37 & 0.81 & 0.56 & 2.69 & 0.92\\
\midrule
\multirow{3}{*}{Qwen2.5-3B} & zero-shot & 60.17 & 0.62 & 0.84 & 0.64 & 2.56 & 0.85\\
                      & one-shot & 54.28 & 0.54 & 0.82 & 0.57 & 2.27 & 0.86 \\
                      & three-shot & 60.45 & 0.63 & 0.84 & 0.59 & 2.12 & 0.86\\
\midrule
\multirow{3}{*}{Qwen2.5-7B} & zero-shot & 63.12 & 0.65 & 0.85 & 0.60 & 1.44 & 0.99\\
                      & one-shot & 65.01 & 0.69 & 0.84 & 0.61 & 1.68 & 1.00\\
                      & three-shot & 65.84 & 0.70 & 0.84 & 0.62 & 1.64 & 1.00\\
\midrule
\multirow{3}{*}{Qwen2.5-14B} & zero-shot & 68.34 & 0.75 & 0.84 & 0.61 & 1.84 & 0.99\\
                      & one-shot & 68.45 & 0.75 & 0.84 & 0.61 & 1.77 & 1.00\\
                      & three-shot & 68.39 & 0.75 & 0.84 & 0.61 & 1.81 & 1.00\\
\midrule
\multirow{3}{*}{Qwen2.5-32B} & zero-shot & 68.74 & 0.76 & 0.83 & 0.61 & 1.87 & 1.00\\
                      & one-shot & 69.05 & 0.76 & 0.84 & 0.61 & 1.77 & 1.00\\
                      & three-shot & 68.82 & 0.76 & 0.83 & 0.61 & 1.82 & 1.00\\
\midrule
\multirow{3}{*}{Qwen2.5-72B} & zero-shot & 69.99 & 0.78 & 0.83 & 0.60 & 1.75 & 1.00\\
                      & one-shot & 69.48 &0.77 & 0.83 & 0.60 & 1.70 & 1.00\\
                      & three-shot & \textbf{71.30} & 0.80 & 0.83 & 0.60 & 1.69 & 1.00\\
\midrule
\multirow{3}{*}{Llama3.1-8B} & zero-shot & 37.02 & 0.30 & 0.74 & 0.51 & 3.60 & 0.88\\
                      & one-shot & 34.54 & 0.28 & 0.68 & 0.49 & 3.97 & 0.92\\
                      & three-shot & 32.45 & 0.27 & 0.61 & 0.46 & 3.89 & 0.93\\
\midrule
\multirow{3}{*}{Llama3.1-70B} & zero-shot & 41.56 & 0.35 & 0.76 & 0.54 & 2.24 & 0.57\\
                      & one-shot & 52.28 & 0.50 & 0.81 & 0.60 & 2.18 & 0.80\\
                      & three-shot & 51.02 & 0.48 & 0.81 & 0.61 & 2.39 & 0.90\\
\midrule
\bottomrule
\end{tabular}
\caption{Open Source Model Results with GPT-4o Evaluation.}
\label{table:llm eval}
\end{threeparttable}
\end{table}

\begin{table}[ht]
\centering
\begin{threeparttable}
\begin{tabular}{c|c|cccccc}
\toprule
Model & Few-shot & $S_{final}$ & $S_{co}$ & $S_{rele}$ & $S_{simi}$ & $S_c$  & Pass rate\\
\midrule
\midrule
\multicolumn{8}{c}{Close Source Model} \\
\hline
\midrule
Deepseek-V2.5 \tnote{*} & three-shot & \textbf{73.14} & 0.81 & 0.86 & 0.63 & 1.52 & 0.99\\
\midrule
ERNIE-3.5 & three-shot & 72.19 & 0.80 & 0.87 & 0.64 & 1.87 & 1.00 \\
\midrule
Moonshoot-v1 & three-shot & 70.89 & 0.77 & 0.87 & 0.64 & 1.59 & 0.99 \\
\midrule
GPT-4o & three-shot & 71.33 & 0.79 & 0.85 & 0.62 & 1.67 & 1.00\\
\bottomrule
\end{tabular}
\begin{tablenotes} 
\footnotesize
\item[*] This model is open-source, but due to its large parameter volume, we call it using the API.
\end{tablenotes} 
\caption{Open Source Model Results with Qwen2.5-72B Evaluation.}
\label{table:llm eval 2}
\end{threeparttable}
\end{table}


We utilized 14 models to operate Level-Navi Agent, encompassing open-source and closed-source models. \textbf{a) Open-source.} The open-source models we used primarily come from the Chinese community, including InternLM series (\cite{cai2024internlm2}), GLM-4 (\cite{glm2024chatglm}), Qwen series (\cite{qwen}) and Llama series (\cite{llama3modelcard}). \textbf{b) Closed-source} For closed-source models, we utilized ERNIE-3.5 from Baidu, Moonshot-v1 from Moonshot AI, and GPT-4o (\cite{achiam2023gpt}) from OpenAI. Note that Deepseek-V2.5 (\cite{deepseekv2}) is an open-source model, but due to its large parameter size, we call it in the form of an API. For the convenience of comparison, we have categorized it in the same group as other closed-source models that are also experimented with using the API.

\subsection{Experimental Results of Our Agent on the Web24 Dataset}\label{experiments result}

Table \ref{table:llm eval} and Table \ref{table:llm eval 2} presents the experimental results obtained by LLMs under our Level-Navi Agent framework and benchmark testing. From the data in the table, we can observe that Qwen2.5-72B and Deepseek-V2.5 performs the best. After summarizing and analyzing all the experiment results, we have analyzed and summarized several key points about the Web Search Agent:

\textbf{Diminishing Marginal Returns of Model Parameters.} From the results obtained from the open-source models of the Internlm series, Qwen series, and Lllama series in Table \ref{table:llm eval}, we observed an empirical pattern: within the same series of models, the larger the parameter count, the higher the final score. Focusing on the scores of the Qwen series models, we also observed the diminishing marginal returns of model parameter size. From 3B to 14B, an approximately fivefold increase in model size led to a performance improvement of about 6 points; from 14B to 72B, a similar fivefold increase in parameter size resulted in only a 3-point improvement. In the results of closed-source models in Table \ref{table:llm eval 2}, we found that although the performance of closed-source models is quite good, Deepseek-V2.5 actually achieves the best among them, which is surprising considering that it is actually an open-source model called in the form of an API. Considering the parameter differences between Qwen2.5-72B and Deepseek-V2.5, this result is acceptable. From the analysis above, it can be seen that: To further enhance the performance of LLMs in executing web search tasks, researchers should focus on how to filter and obtain higher quality information sources, since that the diminishing marginal effect of model parameter is quite clear.

\textbf{Few-shot Prompts Enhance Pass Rates.} For all models, we implemented three types of prompt methods: zero-shot, one-shot, and three-shot (\cite{10.5555/3495724.3495883}). From the perspective of the model's scoring, it seems that different prompting methods do not have a significant impact on the model. Therefore, we calculated the pass rate to reflect the effect of different prompting methods. During the execution of tasks by the Agent, factors such as incorrect LLM output formatting and incorrect function calls can lead to the task not being completed. Hence, we calculated the number of errors for each evaluation and compared it with the total number of the dataset to derive the pass rate, reflecting the performance of the model in executing instructions. Chain of thought method and few-shot prompt combination have been proven effective in previous research (\cite{liang2023prompting}, \cite{ma-etal-2023-chain-thought}), this conclusion is also reflected in our experiments. From Table \ref{table:llm eval}, it can be seen that the three-shot method significantly improved the pass rate of the Agent compared to the zero-shot approach. For some models, such as the Qwen series models with more than 7B parameters, the pass rate is close to 1 even under zero-shot conditions, which also reflects the superior performance of the models themselves. In general, we recommend providing few-shot prompts when executing agent tasks. This approach is not only simple and cost-effective, but also enhances the model's performance in various aspects.

\textbf{Native Chinese LLM Outperform in Chinese Context Tasks.} In our test results, the performance of the Llama series was not very satisfactory, which was reflected in both the final scores and the pass rates. We speculate that this is because our Agent framework and dataset are built around Chinese text (\cite{yuan2023multilingual}, \cite{zhang2023don}). For LLMs like Llama, whose primary application scenarios are in English, this may result in a loss of performance, further elaboration on this aspect will be provided in Section \ref{task fidelity}. In earlier experiments, we tested other well-known open-source LLMs from the English community such as Gemma (\cite{team2024gemma}) and Mistral, but found that their fidelity to Chinese instructions was very poor—they either failed to follow Chinese instructions or were unable to respond to users in Chinese. This finding highlights the importance of multilingual optimization for LLMs. At the same time, our experiments demonstrate, to some extent, the true capabilities and advantages of native Chinese LLMs in Chinese contexts.

\subsection{Comparison with Other Products}

\begin{figure}[htbp!]
    \centering    
    \includegraphics[width=0.8\columnwidth]{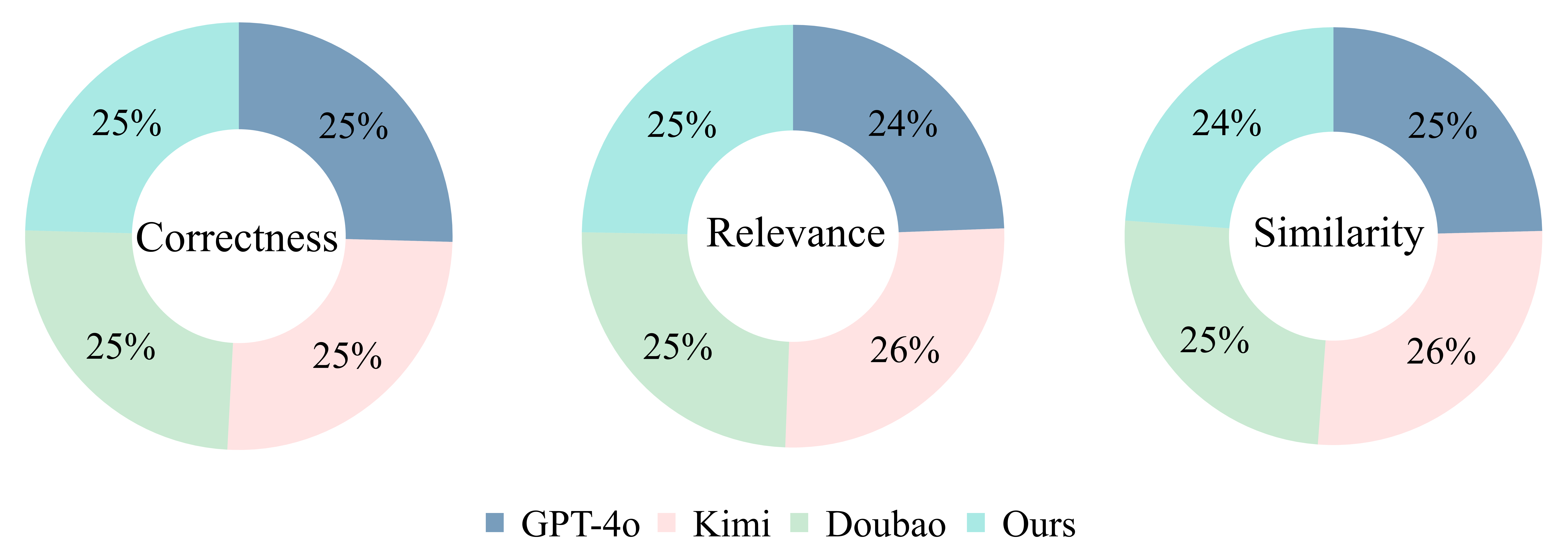}
    \caption{Comparison with other products based on our metrics.}
    \label{fig:compare}
\end{figure}

We also compared our Agent with mature products on the market. We selected two well-known Chinese LLM service providers: Kimi\footnote{\url{https://kimi.moonshot.cn}} from Moonshot AI and Doubao\footnote{\url{https://www.doubao.com}} from ByteDance, while also including the renowned GPT-4o from OpenAI in the comparison. It is noted that all the above products have the ability to conduct web search. We randomly selected 100 examples from the web24 dataset to obtain answers on the product website, and our Agent used Qwen2.5-72b as the execution model.

As depicted in Fig. \ref{fig:compare}, we compared the correctness, relevance, and semantic similarity scores of four objects using a circular diagram. It can be seen from the Fig. \ref{fig:compare} that there is not much difference among the three products provided by LLM service supplier. Kimi performs slightly better than the other products in three metrics, but there is no significant gap. Our Agent has reached the same level as commercial products in these three metrics, which is sufficient to prove the good performance of our framework. At the same time, any user can switch the model they use according to their own situation at any time, making it more flexible and cost-effective.

\subsection{Analysis of Metrics}\label{traditional metrics}

\noindent \textbf{Limitation of Traditional Metrics.} In Section \ref{evaluation method}, We briefly discussed the limitations of traditional evaluation methods, and next we will further elaborate on this point through detailed experiments and analysis. 

Table \ref{table:token} presents the results of evaluating model responses using traditional methods. We use a Chinese tokenizer to tokenize the model's responses and the ground truth answers, and then calculate the recall and F1 scores using statistical methods. By comparing and analyzing the performance of LLMs with different parameter sizes on the same task, we discovered a counterintuitive phenomenon: the increase in model parameters did not universally lead to an improvement in F1 scores; instead, in some cases, we observed a decline in F1 scores. Concurrently, the recall scores exhibited a clear upward trend with the increase in model parameters. 

After examining the analysis model's response, we discovered the cause of this phenomenon: Web search is a complex and open-ended task. Although the answers we label already include the correct key information, the LLM might generate more comprehensive information related to the question according to the context in the end. For the LLM, doing so enriches the depth and credibility of the answer. However, in terms of F1 score, longer texts are more likely to lead to a mismatch between the score and the original answer, resulting in a decrease in the score. From the perspective of recall evaluation, the longer the relevant text generated by the LLM, the higher the score will be, assuming the ground truth answer remains unchanged. This is also reflected in Table \ref{table:token}. Similarly, for ROUGE score evaluation (\cite{lin2004rouge}), the aforementioned reasons also prevent us from obtaining an accurate reflection of the LLM's capabilities from the scores. These phenomena all demonstrate the limitations of traditional metrics.

\noindent \textbf{The effectiveness of our metrics.} From Tables \ref{table:llm eval} and \ref{table:llm eval 2}, it can be observed that using our metrics, the performance distribution of various models aligns with empirical knowledge and common sense. In terms of Correctness Scores, the advantage brought by model parameter size is clearly demonstrated. Meanwhile, Semantic Similarity Scores and Relevance Scores consistently reflect the capability differences among models. Through the overall scores, anyone can intuitively discern the performance differences between models. These findings strongly validate the effectiveness of our metrics.

Therefore, we believe that traditional token-based evaluation methods cannot accurately reflect the quality of model responses when dealing with web search Q\&A tasks that involve summarizing diverse information. At present, it seems that using human evaluation or LLM evaluation can better assess model responses that include rich relevant materials and different semantic expressions(\cite{zhuge2024agent}).
\begin{table}[ht]
\centering
\begin{tabular}{l|ccc}
\toprule
Model &ROUGE & F1 & Recall \\
\midrule
Internlm2.5-7B  & 0.12 & 0.09 & 0.69 \\
Internlm2.5-20B & 0.12 & 0.09 & 0.74 \\
\midrule
Qwen2.5-3B & 0.2 & 0.18 & 0.71 \\
Qwen2.5-7B  & 0.23 & 0.21 & 0.74 \\
Qwen2.5-14B & 0.19 & 0.16 & 0.78 \\
Qwen2.5-32B & 0.19 & 0.16 & 0.78 \\
Qwen2.5-72B & 0.16 & 0.12 & 0.81 \\
\bottomrule
\end{tabular}
\caption{Using traditional metrics (Models are under the three-shot method).}
\label{table:token}
\end{table}

\subsection{Error Analysis and Discussion}\label{error analysis}

In this section, we will conduct an error analysis through experimental data to identify the causes of poor model performance, and also provide corresponding suggestions on how to improve model performance.

\subsubsection{Overconfidence Phenomenon in Web Search Function Usage.}

\begin{figure}[htbp!]
    \centering    
    \includegraphics[width=0.9\columnwidth]{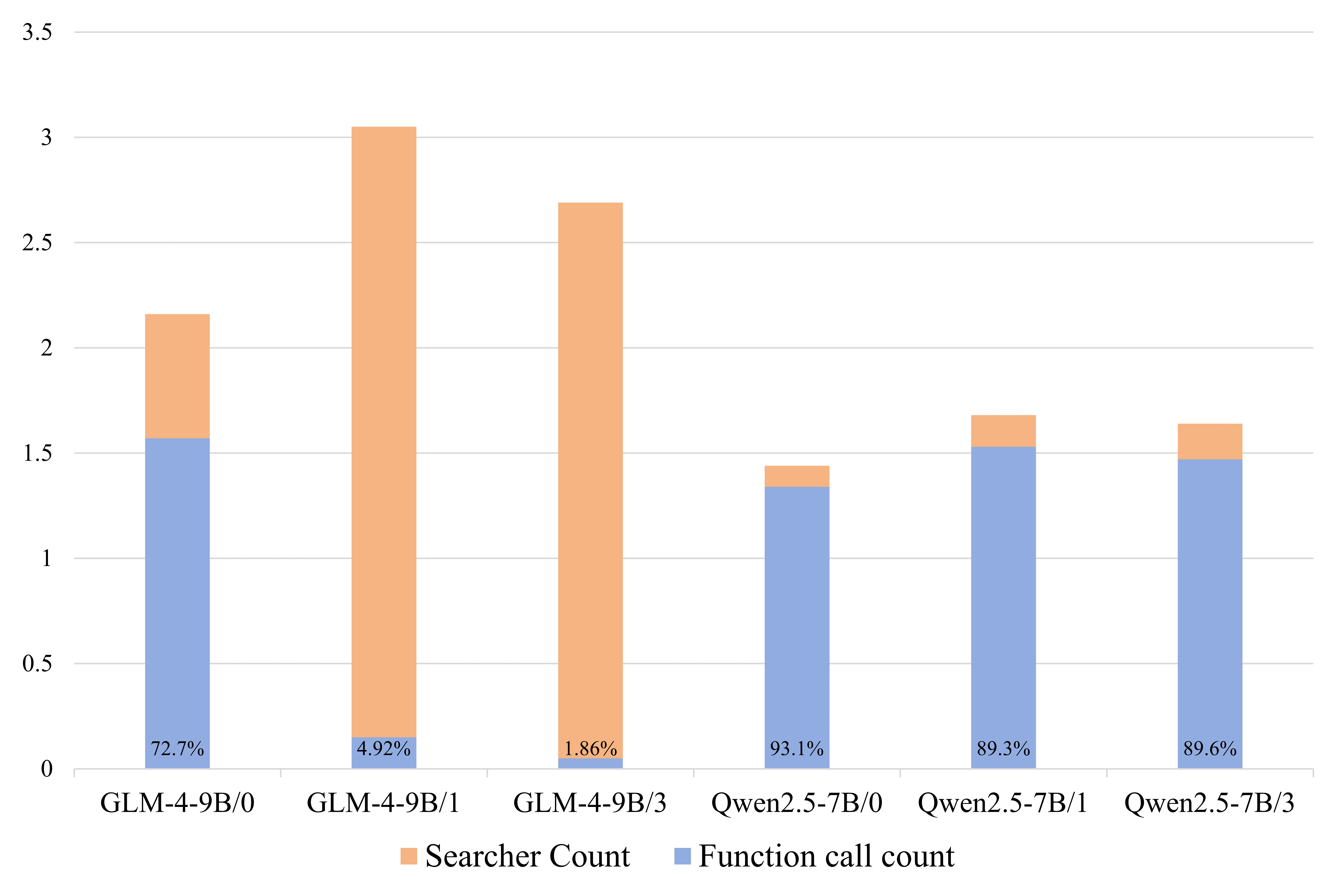}
    \caption{Comparison of Searcher and Function call counts, the percentage at the bottom represent the Searcher count / Function call count(0 for zero-shot; 1 for one-shot; 3 for three-shot).}
    \label{fig:error1}
\end{figure}

In Table \ref{table:llm eval}, the scores of GLM-4-9B are lower than expected. We uncovered the reason for this phenomenon by calculating the average Search Count invoked per task and the actual number of Web Search Function calls. In Fig. \ref{fig:error1}, we compared the gap between GLM-4 and Qwen2.5-7B on the aforementioned metrics. We observed that Qwen2.5-7B's function call rate reached about 90\% of the Agent invocations, while GLM-4-9B's ratio dropped from 30\% to a single-digit percentage. Given that 70\% of the answers in the dataset come from news sources, GLM-4-9B cannot possibly answer correctly without invoking web searches. We refer to this phenomenon as ``Overconfidence"\cite{10.1145/3703155}.

The issue indicates that LLMs might overestimate their question-answering abilities during training, overlooking the need for external resources in certain situations \cite{xiongcan}. To address this overconfidence, we recommend that developers balance positive and negative examples in the training dataset to improve LLMs' function-calling capabilities.

\subsubsection{Low Task Fidelity in Conducting Chinese Tasks.}\label{task fidelity}

\begin{table}[ht]
\centering
\begin{tabular}{l|cc}
\toprule
Model & Few-shot & Non-compliant Response(\%)  \\
\midrule
\multirow{3}{*}{Llama3.1-8B} & zero-shot & 25.16 \\
                      & one-shot &  27.86\\
                      & three-shot &  23.64\\
\midrule
\multirow{3}{*}{Llama3.1-70B} & zero-shot & 2.08\\
                      & one-shot & 0.42\\
                      & three-shot &  0.42\\
\bottomrule
\end{tabular}
\caption{Statistics of non-compliant responses generated by the Llama 3.1 series models.}
\label{table:error2}
\end{table}

When assessing LLM agents, we prioritize whether the model comprehends and adheres to instructions to answer questions. We call it ``Task Fidelity",  which reflects the model's faithfulness in executing instructions. The Relevance Scores do not take into account the correctness of the model's response, so it can reflect the end-to-end Task Fidelity.

In Table \ref{table:llm eval}, the relevance scores of Llama3.1-8B did not behave as expected compared to other models; instead, they fluctuated significantly. Upon examining the outputs of the Llama 3.1 series models, we found that a considerable portion of the responses did not fully comply with our instructions, with some incorrectly mixing the given instructions with the answers. Table \ref{table:error2} more detailedly reflects this type of non-compliant response. The introduction of the few-shot method did not improve this issue for Llama3.1-8B, only Llama3.1-70B showed improvement.

We view non-compliant responses as indicating low Task Fidelity. The LLM's struggle to grasp the intent of instructions in lengthy Chinese contexts can greatly impair task performance. Developers should focus on ensuring smaller LLMs to maintain the multilingual ability as bigger ones.
\section{Conclusion}

In this paper, we introduce the Level-Navi Agent and a novel benchmark for web search tasks. The Level-Navi Agent offers an innovative solution to web search challenges through the collaboration of multiple agents and a hierarchical approach to reasoning and searching. Starting from the Chinese open-source community, we employed new evaluation metrics and methods to comprehensively assess the performance of various LLMs in executing web search tasks. This analysis sheds light on the true capabilities of current LLMs when performing web search tasks within the Chinese Internet. Through data-driven error analysis, we identify the limitations of LLMs in handling web search tasks and provide recommendations for improvement, contributing to the advancement of this field.



\bibliography{reference}

\begin{thebibliography}{36}
\providecommand{\natexlab}[1]{#1}
\providecommand{\url}[1]{\texttt{#1}}
\expandafter\ifx\csname urlstyle\endcsname\relax
  \providecommand{\doi}[1]{doi: #1}\else
  \providecommand{\doi}{doi: \begingroup \urlstyle{rm}\Url}\fi

\bibitem[Achiam et~al.(2023)Achiam, Adler, Agarwal, Ahmad, Akkaya, Aleman, Almeida, Altenschmidt, Altman, Anadkat, et~al.]{achiam2023gpt}
Josh Achiam, Steven Adler, Sandhini Agarwal, Lama Ahmad, Ilge Akkaya, Florencia~Leoni Aleman, Diogo Almeida, Janko Altenschmidt, Sam Altman, Shyamal Anadkat, et~al.
\newblock Gpt-4 technical report.
\newblock \emph{arXiv preprint arXiv:2303.08774}, 2023.

\bibitem[AI@Meta(2024)]{llama3modelcard}
AI@Meta.
\newblock Llama 3 model card.
\newblock 2024.
\newblock URL \url{https://github.com/meta-llama/llama3/blob/main/MODEL_CARD.md}.

\bibitem[Asai et~al.(2023)Asai, Wu, Wang, Sil, and Hajishirzi]{asai2023self}
Akari Asai, Zeqiu Wu, Yizhong Wang, Avirup Sil, and Hannaneh Hajishirzi.
\newblock Self-rag: Self-reflective retrieval augmented generation.
\newblock In \emph{NeurIPS 2023 Workshop on Instruction Tuning and Instruction Following}, 2023.

\bibitem[Bai et~al.(2023)Bai, Bai, Chu, Cui, Dang, Deng, Fan, Ge, Han, Huang, Hui, Ji, and et~al.]{qwen}
Jinze Bai, Shuai Bai, Yunfei Chu, Zeyu Cui, Kai Dang, Xiaodong Deng, Yang Fan, Wenbin Ge, Yu~Han, Fei Huang, Binyuan Hui, Luo Ji, and Mei~Li et~al.
\newblock Qwen technical report.
\newblock \emph{arXiv preprint arXiv:2309.16609}, 2023.

\bibitem[Brin \& Page(1998)Brin and Page]{BRIN1998107}
Sergey Brin and Lawrence Page.
\newblock The anatomy of a large-scale hypertextual web search engine.
\newblock \emph{Computer Networks and ISDN Systems}, 30\penalty0 (1):\penalty0 107--117, 1998.
\newblock ISSN 0169-7552.
\newblock \doi{https://doi.org/10.1016/S0169-7552(98)00110-X}.
\newblock URL \url{https://www.sciencedirect.com/science/article/pii/S016975529800110X}.
\newblock Proceedings of the Seventh International World Wide Web Conference.

\bibitem[Brown et~al.(2020)Brown, Mann, Ryder, Subbiah, Kaplan, Dhariwal, Neelakantan, Shyam, Sastry, Askell, and et~al.]{10.5555/3495724.3495883}
Tom~B. Brown, Benjamin Mann, Nick Ryder, Melanie Subbiah, Jared Kaplan, Prafulla Dhariwal, Arvind Neelakantan, Pranav Shyam, Girish Sastry, Amanda Askell, and Agarwal et~al.
\newblock Language models are few-shot learners.
\newblock In \emph{Proceedings of the 34th International Conference on Neural Information Processing Systems}, NIPS '20, Red Hook, NY, USA, 2020. Curran Associates Inc.
\newblock ISBN 9781713829546.

\bibitem[Cai et~al.(2024)Cai, Cao, Chen, Chen, Chen, Chen, Chen, Chen, Chen, Chu, Dong, Duan, Fan, Fei, Gao, Ge, Gu, Gu, Gui, Guo, Guo, and et~al.]{cai2024internlm2}
Zheng Cai, Maosong Cao, Haojiong Chen, Kai Chen, Keyu Chen, Xin Chen, Xun Chen, Zehui Chen, Zhi Chen, Pei Chu, Xiaoyi Dong, Haodong Duan, Qi~Fan, Zhaoye Fei, Yang Gao, Jiaye Ge, Chenya Gu, Yuzhe Gu, Tao Gui, Aijia Guo, Qipeng Guo, and Conghui~He et~al.
\newblock Internlm2 technical report, 2024.

\bibitem[Chan et~al.(2024)Chan, Xu, Yuan, Luo, Xue, Guo, and Fu]{chan2024rqrag}
Chi-Min Chan, Chunpu Xu, Ruibin Yuan, Hongyin Luo, Wei Xue, Yike Guo, and Jie Fu.
\newblock {RQ}-{RAG}: Learning to refine queries for retrieval augmented generation.
\newblock In \emph{First Conference on Language Modeling}, 2024.
\newblock URL \url{https://openreview.net/forum?id=tzE7VqsaJ4}.

\bibitem[Chen et~al.(2024)Chen, Liu, Wang, Liu, Zhang, Chen, and Zhao]{chen2024mindsearch}
Zehui Chen, Kuikun Liu, Qiuchen Wang, Jiangning Liu, Wenwei Zhang, Kai Chen, and Feng Zhao.
\newblock Mindsearch: Mimicking human minds elicits deep ai searcher.
\newblock \emph{arXiv preprint arXiv:2407.20183}, 2024.

\bibitem[DeepSeek-AI(2024)]{deepseekv2}
DeepSeek-AI.
\newblock Deepseek-v2: A strong, economical, and efficient mixture-of-experts language model, 2024.

\bibitem[Es et~al.(2024)Es, James, Espinosa~Anke, and Schockaert]{es-etal-2024-ragas}
Shahul Es, Jithin James, Luis Espinosa~Anke, and Steven Schockaert.
\newblock {RAGA}s: Automated evaluation of retrieval augmented generation.
\newblock In Nikolaos Aletras and Orphee De~Clercq (eds.), \emph{Proceedings of the 18th Conference of the European Chapter of the Association for Computational Linguistics: System Demonstrations}, pp.\  150--158, St. Julians, Malta, March 2024. Association for Computational Linguistics.
\newblock URL \url{https://aclanthology.org/2024.eacl-demo.16}.

\bibitem[GLM et~al.(2024)GLM, Zeng, Xu, Wang, Zhang, Yin, Rojas, Feng, Zhao, Lai, Yu, Wang, Sun, and et~al.]{glm2024chatglm}
Team GLM, Aohan Zeng, Bin Xu, Bowen Wang, Chenhui Zhang, Da~Yin, Diego Rojas, Guanyu Feng, Hanlin Zhao, Hanyu Lai, Hao Yu, Hongning Wang, Jiadai Sun, and Jiajie~Zhang et~al.
\newblock Chatglm: A family of large language models from glm-130b to glm-4 all tools, 2024.

\bibitem[Huang et~al.(2024)Huang, Yu, Ma, Zhong, Feng, Wang, Chen, Peng, Feng, Qin, and Liu]{10.1145/3703155}
Lei Huang, Weijiang Yu, Weitao Ma, Weihong Zhong, Zhangyin Feng, Haotian Wang, Qianglong Chen, Weihua Peng, Xiaocheng Feng, Bing Qin, and Ting Liu.
\newblock A survey on hallucination in large language models: Principles, taxonomy, challenges, and open questions.
\newblock \emph{ACM Trans. Inf. Syst.}, November 2024.
\newblock ISSN 1046-8188.
\newblock \doi{10.1145/3703155}.
\newblock URL \url{https://doi.org/10.1145/3703155}.

\bibitem[Jiang et~al.(2024)Jiang, Zhang, Guo, Wu, Lei, Qiu, Lu, Chen, Song, Gao, et~al.]{jiang2024mmsearch}
Dongzhi Jiang, Renrui Zhang, Ziyu Guo, Yanmin Wu, Jiayi Lei, Pengshuo Qiu, Pan Lu, Zehui Chen, Guanglu Song, Peng Gao, et~al.
\newblock Mmsearch: Benchmarking the potential of large models as multi-modal search engines.
\newblock \emph{arXiv preprint arXiv:2409.12959}, 2024.

\bibitem[Jin et~al.(2024)Jin, Yu, Shu, Zhao, Hua, Meng, Zhang, and Du]{jin2024impact}
Mingyu Jin, Qinkai Yu, Dong Shu, Haiyan Zhao, Wenyue Hua, Yanda Meng, Yongfeng Zhang, and Mengnan Du.
\newblock The impact of reasoning step length on large language models.
\newblock \emph{arXiv preprint arXiv:2401.04925}, 2024.

\bibitem[Li et~al.(2023)Li, Zhang, Koto, Yang, Zhao, Gong, Duan, and Baldwin]{li2023cmmlu}
Haonan Li, Yixuan Zhang, Fajri Koto, Yifei Yang, Hai Zhao, Yeyun Gong, Nan Duan, and Timothy Baldwin.
\newblock Cmmlu: Measuring massive multitask language understanding in chinese.
\newblock \emph{arXiv preprint arXiv:2306.09212}, 2023.

\bibitem[Liang et~al.(2023)Liang, Wang, Zhu, Wang, Qian, and Lan]{liang2023prompting}
Yuanyuan Liang, Jianing Wang, Hanlun Zhu, Lei Wang, Weining Qian, and Yunshi Lan.
\newblock Prompting large language models with chain-of-thought for few-shot knowledge base question generation.
\newblock \emph{arXiv preprint arXiv:2310.08395}, 2023.

\bibitem[Lin(2004)]{lin2004rouge}
Chin-Yew Lin.
\newblock Rouge: A package for automatic evaluation of summaries.
\newblock In \emph{Text summarization branches out}, pp.\  74--81, 2004.

\bibitem[Liu et~al.(2024)Liu, Lei, Wang, Huang, Feng, Wen, Cheng, Ke, Xu, Tam, Zhang, Sun, Gu, Wang, Zhang, Huang, Dong, and Tang]{liu-etal-2024-alignbench}
Xiao Liu, Xuanyu Lei, Shengyuan Wang, Yue Huang, Andrew Feng, Bosi Wen, Jiale Cheng, Pei Ke, Yifan Xu, Weng~Lam Tam, Xiaohan Zhang, Lichao Sun, Xiaotao Gu, Hongning Wang, Jing Zhang, Minlie Huang, Yuxiao Dong, and Jie Tang.
\newblock {A}lign{B}ench: Benchmarking {C}hinese alignment of large language models.
\newblock In Lun-Wei Ku, Andre Martins, and Vivek Srikumar (eds.), \emph{Proceedings of the 62nd Annual Meeting of the Association for Computational Linguistics (Volume 1: Long Papers)}, pp.\  11621--11640, Bangkok, Thailand, August 2024. Association for Computational Linguistics.
\newblock \doi{10.18653/v1/2024.acl-long.624}.
\newblock URL \url{https://aclanthology.org/2024.acl-long.624}.

\bibitem[Ma et~al.(2023)Ma, Li, and Zhang]{ma-etal-2023-chain-thought}
Xilai Ma, Jing Li, and Min Zhang.
\newblock Chain of thought with explicit evidence reasoning for few-shot relation extraction.
\newblock In Houda Bouamor, Juan Pino, and Kalika Bali (eds.), \emph{Findings of the Association for Computational Linguistics: EMNLP 2023}, pp.\  2334--2352, Singapore, December 2023. Association for Computational Linguistics.
\newblock \doi{10.18653/v1/2023.findings-emnlp.153}.
\newblock URL \url{https://aclanthology.org/2023.findings-emnlp.153}.

\bibitem[Nakano et~al.(2021)Nakano, Hilton, Balaji, Wu, Long, Kim, Hesse, Jain, Kosaraju, Saunders, Jiang, Cobbe, Eloundou, Krueger, Button, Knight, Chess, and Schulman]{Nakano2021WebGPTBQ}
Reiichiro Nakano, Jacob Hilton, Suchir Balaji, Jeff Wu, Ouyang Long, Christina Kim, Christopher Hesse, Shantanu Jain, Vineet Kosaraju, William Saunders, Xu~Jiang, Karl Cobbe, Tyna Eloundou, Gretchen Krueger, Kevin Button, Matthew Knight, Benjamin Chess, and John Schulman.
\newblock Webgpt: Browser-assisted question-answering with human feedback.
\newblock \emph{ArXiv}, abs/2112.09332, 2021.
\newblock URL \url{https://api.semanticscholar.org/CorpusID:245329531}.

\bibitem[Ram et~al.(2023)Ram, Levine, Dalmedigos, Muhlgay, Shashua, Leyton-Brown, and Shoham]{ram-etal-2023-context}
Ori Ram, Yoav Levine, Itay Dalmedigos, Dor Muhlgay, Amnon Shashua, Kevin Leyton-Brown, and Yoav Shoham.
\newblock In-context retrieval-augmented language models.
\newblock \emph{Transactions of the Association for Computational Linguistics}, 11:\penalty0 1316--1331, 2023.
\newblock \doi{10.1162/tacl_a_00605}.
\newblock URL \url{https://aclanthology.org/2023.tacl-1.75}.

\bibitem[Reddy et~al.(2023)Reddy, Fung, Zeng, Li, Wang, Sullivan, and Ji]{Reddy2023SmartBookAS}
Revanth~Gangi Reddy, Yi~Ren Fung, Qi~Zeng, Manling Li, Ziqi Wang, Paul Sullivan, and Heng Ji.
\newblock Smartbook: Ai-assisted situation report generation.
\newblock \emph{ArXiv}, abs/2303.14337, 2023.
\newblock URL \url{https://api.semanticscholar.org/CorpusID:257766360}.

\bibitem[Reddy et~al.(2024)Reddy, Mukherjee, Kim, Wang, Hakkani-Tur, and Ji]{reddy2024infogent}
Revanth~Gangi Reddy, Sagnik Mukherjee, Jeonghwan Kim, Zhenhailong Wang, Dilek Hakkani-Tur, and Heng Ji.
\newblock Infogent: An agent-based framework for web information aggregation.
\newblock \emph{arXiv preprint arXiv:2410.19054}, 2024.

\bibitem[Siriwardhana et~al.(2023)Siriwardhana, Weerasekera, Wen, Kaluarachchi, Rana, and Nanayakkara]{siriwardhana2023improving}
Shamane Siriwardhana, Rivindu Weerasekera, Elliott Wen, Tharindu Kaluarachchi, Rajib Rana, and Suranga Nanayakkara.
\newblock Improving the domain adaptation of retrieval augmented generation (rag) models for open domain question answering.
\newblock \emph{Transactions of the Association for Computational Linguistics}, 11:\penalty0 1--17, 2023.

\bibitem[Spatharioti et~al.(2023)Spatharioti, Rothschild, Goldstein, and Hofman]{spatharioti2023comparing}
Sofia~Eleni Spatharioti, David~M Rothschild, Daniel~G Goldstein, and Jake~M Hofman.
\newblock Comparing traditional and llm-based search for consumer choice: A randomized experiment.
\newblock \emph{arXiv preprint arXiv:2307.03744}, 2023.

\bibitem[Team et~al.(2024)Team, Riviere, Pathak, Sessa, Hardin, Bhupatiraju, Hussenot, Mesnard, Shahriari, Ram{\'e}, et~al.]{team2024gemma}
Gemma Team, Morgane Riviere, Shreya Pathak, Pier~Giuseppe Sessa, Cassidy Hardin, Surya Bhupatiraju, L{\'e}onard Hussenot, Thomas Mesnard, Bobak Shahriari, Alexandre Ram{\'e}, et~al.
\newblock Gemma 2: Improving open language models at a practical size.
\newblock \emph{arXiv preprint arXiv:2408.00118}, 2024.

\bibitem[Wang et~al.(2023)Wang, Duan, Wang, Li, Xian, Yin, Rong, and Xiong]{wang2023knowledge}
Keheng Wang, Feiyu Duan, Sirui Wang, Peiguang Li, Yunsen Xian, Chuantao Yin, Wenge Rong, and Zhang Xiong.
\newblock Knowledge-driven cot: Exploring faithful reasoning in llms for knowledge-intensive question answering.
\newblock \emph{arXiv preprint arXiv:2308.13259}, 2023.

\bibitem[Wei et~al.(2022)Wei, Wang, Schuurmans, Bosma, Xia, Chi, Le, Zhou, et~al.]{wei2022chain}
Jason Wei, Xuezhi Wang, Dale Schuurmans, Maarten Bosma, Fei Xia, Ed~Chi, Quoc~V Le, Denny Zhou, et~al.
\newblock Chain-of-thought prompting elicits reasoning in large language models.
\newblock \emph{Advances in neural information processing systems}, 35:\penalty0 24824--24837, 2022.

\bibitem[Xiao et~al.(2024)Xiao, Liu, Zhang, Muennighoff, Lian, and Nie]{xiao2024c}
Shitao Xiao, Zheng Liu, Peitian Zhang, Niklas Muennighoff, Defu Lian, and Jian-Yun Nie.
\newblock C-pack: Packed resources for general chinese embeddings.
\newblock In \emph{Proceedings of the 47th International ACM SIGIR Conference on Research and Development in Information Retrieval}, pp.\  641--649, 2024.

\bibitem[Xiong et~al.(2024)Xiong, Hu, Lu, LI, Fu, He, and Hooi]{xiongcan}
Miao Xiong, Zhiyuan Hu, Xinyang Lu, YIFEI LI, Jie Fu, Junxian He, and Bryan Hooi.
\newblock Can llms express their uncertainty? an empirical evaluation of confidence elicitation in llms.
\newblock In \emph{The Twelfth International Conference on Learning Representations}, 2024.

\bibitem[Yang et~al.(2024)Yang, Sun, Xin, Sun, Bhalla, Chen, Choudhary, Gui, Jiang, Jiang, et~al.]{yang2024crag}
Xiao Yang, Kai Sun, Hao Xin, Yushi Sun, Nikita Bhalla, Xiangsen Chen, Sajal Choudhary, Rongze~Daniel Gui, Ziran~Will Jiang, Ziyu Jiang, et~al.
\newblock Crag--comprehensive rag benchmark.
\newblock \emph{arXiv preprint arXiv:2406.04744}, 2024.

\bibitem[Ye et~al.(2023)Ye, Chen, Xu, Zu, Shao, Liu, Cui, Zhou, Gong, Shen, et~al.]{ye2023comprehensive}
Junjie Ye, Xuanting Chen, Nuo Xu, Can Zu, Zekai Shao, Shichun Liu, Yuhan Cui, Zeyang Zhou, Chao Gong, Yang Shen, et~al.
\newblock A comprehensive capability analysis of gpt-3 and gpt-3.5 series models.
\newblock \emph{arXiv preprint arXiv:2303.10420}, 2023.

\bibitem[Yuan et~al.(2023)Yuan, Yuan, Wu, and Li]{yuan2023multilingual}
Fei Yuan, Shuai Yuan, Zhiyong Wu, and Lei Li.
\newblock How multilingual is multilingual llm?
\newblock \emph{arXiv preprint arXiv:2311.09071}, 2023.

\bibitem[Zhang et~al.(2023)Zhang, Li, Hauer, Shi, and Kondrak]{zhang2023don}
Xiang Zhang, Senyu Li, Bradley Hauer, Ning Shi, and Grzegorz Kondrak.
\newblock Don’t trust chatgpt when your question is not in english: A study of multilingual abilities and types of llms.
\newblock In \emph{Proceedings of the 2023 Conference on Empirical Methods in Natural Language Processing}, pp.\  7915--7927, 2023.

\bibitem[Zhuge et~al.(2024)Zhuge, Zhao, Ashley, Wang, Khizbullin, Xiong, Liu, Chang, Krishnamoorthi, Tian, et~al.]{zhuge2024agent}
Mingchen Zhuge, Changsheng Zhao, Dylan Ashley, Wenyi Wang, Dmitrii Khizbullin, Yunyang Xiong, Zechun Liu, Ernie Chang, Raghuraman Krishnamoorthi, Yuandong Tian, et~al.
\newblock Agent-as-a-judge: Evaluate agents with agents.
\newblock \emph{arXiv preprint arXiv:2410.10934}, 2024.

\end{thebibliography}
\bibliographystyle{colm2024_conference}

\end{document}